\documentclass[aps,prd,amsmath,preprint,showpacs,showkeys,eqsecnum]{revtex4}

\usepackage{latexsym}
\usepackage{graphicx}

\begin{document}
\title{Multipole particle in Relativity}

\author{Akira Ohashi}
\email{ohashi@phys.ocha.ac.jp}
\affiliation{Department of Physics, Ochanomizu University,
1-1 Otsuka, 2 Bunkyo, Tokyo, 112-8610, Japan}

\date{\today}

\begin{abstract}
We discuss the motion of extended objects in a spacetime
by considering a gravitational field created by these objects.
We define multipole moments of the objects as
a classification by Lie group $SO(3)$.
Then, we construct an energy-momentum tensor for the objects
and derive equations of motion from it.
As a result, we reproduce the Papapetrou equations
for a spinning particle.
Furthermore, we will show that we can obtain more simple equations
than the Papapetrou equations by changing the center-of-mass.
\end{abstract}

\maketitle

\section{Introduction}
The motion of celestial bodies has attracted people's interests from
the ancient era. The ancient people used to obtain the important information on
the seasons or the time from it.
Even now, the motion of celestial bodies is
one of the important foundations of the modern science.
The modern physics begins with the Kepler's law, the Newton's law and
the law of universal gravitation.
Our understanding about the celestial motion was extended
by the Newton's theory.
The interest in the celestial motion did not change
even when Einstein proposed the general relativity.
Indeed, one of the classical tests of the general relativity
is the account for the anomalous perihelion shift of Mercury
\cite{Will}.
At present, the knowledge of the celestial motion is increasing
its importance as the devices to probe our universe.
For instance, the general relativity predicts
the existence of gravitational waves, and many researchers in the world
endeavor to catch them. One of the main targets in the detection of
gravitational waves is a binary system which consists of celestial bodies
like neutron stars or black holes. Those researchers attempt to detect
the gravitational waves generated by the orbital motion of the binary.
Therefore, it is still important to investigate the motion of
extended bodies like celestial objects in the gravitational field
\cite{Kidder,Tagoshi,Tanaka}.

Now, it is recognized that there are two flows in
the studies on movement of an extended object
in the general theory of relativity.
One of them starts from the study of Einstein, Infeld and Hoffmann \cite{EIH}.
They regarded the existence of object as singular points of
the gravitational field, and discuss the motion of objects
in a weak field approximation.
The advantage of their method is that
it is unnecessary to treat the objects as test particles
because Einstein {\it et.al.} solved the field equations in an approximation.
Their study has led to the later development of
the post Newtonian theory.
Another flow regards the motion of objects as motion of
a test particle on a background spacetime.
In this standpoint, Papapetrou derived
equations of motion for a spinning particle moving
in a curved spacetime by considering the Taylor expansion,
that is, multipole expansion of the divergence of
an energy-momentum tensor at the position of the particle
\cite{Papapetrou}.
The equations derived by him are known as the Papapetrou equations.
The Papapetrou's method was not covariant in
integrating tensors, but Dixon developed a covariant method
by using the bitensor formulation \cite{Dixon}.
In addition, there exists another interesting way to derive
the Papapetrou equations.
Tulczyjew derived the Papapetrou equations in a more simple and covariant way
by using the Dirac's delta function \cite{Tulczyjew}.
More recently, Anandan, Dadhich and Singh developed an approach with
the action principle \cite{ADS}. They consider the Taylor expansion of
an action for a particle and derive equations of motion.
In this article, our standpoint is the one of latter flow.

Now, there are two approaches to treat the multipole properties of objects.
One of them is based on the distribution of the matter, and
another one is based on the gravitational field created by the objects.
In consideration on the motion of objects,
it is indeed intuitive to focus on the distribution of matter
in deriving equations of motion as Papapetrou and Dixon did.
However, when we discuss the motion in the gravitation field,
it is more natural to be based on the field created by the object
than to be based on the distribution of the matter.
In addition, the treatment of multipoles in previous researches was not exactly
based on $SO(3)$, though the concept of multipole moments should be defined by
a classification by the Lie group $SO(3)$.

In this article, our purpose is to discuss the motion of
objects on a curved spacetime based on the above viewpoint.
Our argument will be presented as follows.
In the section \ref{sec:mp in sr},
we classify solutions of the field equations and define multipole moments
based on $SO(3)$ in the context of the special relativity.
Then, we recognize an object as a point singularity of the field
and construct an energy-momentum tensor of the object.
In section \ref{sec:mp in gr}, then we turn to the general relativity
and derive equations of motion from the energy-momentum tensor.
Then, when we examine up to dipole moment,
we show that our equations can be equivalent to the Papapetrou equations
in adopting a certain definition of the center-of-mass.
Moreover, we propose more simple equations of motion for
the spinning particle.
In section \ref{sec:summary}, we summarize our results.

Notations used in this article are the following.
The signature of metrics is $(-,+,+,+)$.
We denote metrics of a Minkowsky spacetime and a curved spacetime
by $\eta_{\mu\nu}$ and $g_{\mu\nu}$, respectively.
The Roman indices run over 1 to 3 and the Greek indices run over
0 to 3. The definition of curvature is
$R^{\mu}{}_{\nu\alpha\beta}=\Gamma^{\mu}{}_{\nu\beta,\alpha}-
\Gamma^{\mu}{}_{\nu\beta,\alpha}+
\Gamma^{\mu}{}_{\alpha\sigma}\Gamma^{\sigma}{}_{\nu\beta}-
\Gamma^{\mu}{}_{\beta\sigma}\Gamma^{\sigma}{}_{\nu\alpha}$ and
$R_{\mu\nu}=R^{\sigma}{}_{\mu\sigma\nu}$.
We use convenient symbols
$\partial_{\alpha\beta\cdots\gamma}=\partial_{\alpha}\partial_{\gamma}
\cdots\partial_{\gamma}$ for simplicity of writing.
We occasionally use the multi-index notation for capital Roman indices;
in the expression $A_LB^L$,
{\it the upper-case letter} "$L$" indicates an ordered set of
{\it lower-case letter} "$l$" {\it indices}, i.e.,
"$l$" is a mere number of the indeces.
For example, $A_LB^L=A_{k_1\cdots k_l}B^{k_1\cdots k_l}$.
The signature of perfect asymmetry symbol $\epsilon$'s is defined
by $\epsilon^{123}=\epsilon_{123}=+1$ for 3-dimensional and
$\epsilon^{0123}={}-\epsilon_{0123}=+1$ for 4-dimensional.

\section{Multipole particle in special relativity}\label{sec:mp in sr}
In this section, we formulate an energy-momentum tensor
which is corresponding to a multipole particle in the inertial motion
on the Minkowsky spacetime.
Here, we suppose that the particle can be a source of gravitation but
its motion receives no influence from the gravitational field.
The interaction between the particle and the gravitational
field will be introduced in the next section.

\subsection{Isolated solution of linearized Einstein equation}
It is well known that
the Einstein equation can be written in
a Klein-Gordon type equation as following,
\begin{eqnarray}\label{eq:KGEinstein}
\Box h^{\mu\nu}=-\frac{16\pi G}{c^4}\left(-g\right)
\left(T^{\mu\nu}+t^{\mu\nu}_{\rm LL}\right)
+\left(h^{\mu\nu}{}_{,\rho\sigma}h^{\rho\sigma}
-h^{\mu\rho}{}_{,\sigma}h^{\nu\sigma}{}_{,\rho}\right),
\end{eqnarray}
where $\Box$ is the d'Alembertian by the Lorenz metric 
$\eta^{\mu\nu}={\rm diag}(-1,1,1,1)$,
$h^{\mu\nu}$ is defined as
$h^{\mu\nu}=\eta^{\mu\nu}-\sqrt{-g}g^{\mu\nu}$,
and $t^{\mu\nu}_{\rm LL}$ is the Landau-Lifshitz's
pseudo-tensor. Moreover,
the de Donder gauge condition $h^{\mu\nu}{}_{,\nu}=0$ is
imposed on $h^{\mu\nu}$ as a coordinate condition.
If we ignore the nonlinear terms in the equation (\ref{eq:KGEinstein}),
we obtain the linearized Einstein equation
\begin{equation}\label{eq:LinearEin}
\Box h^{\mu\nu}=-\frac{16\pi G}{c^4}T^{\mu\nu}.
\end{equation}

The concept of multipole moment for isolated solutions of
the equation (\ref{eq:LinearEin}) is given by a classification based on
the representation of Lie group $SO(3)$ \cite{Thorne}.
When the elements of $SO(3)$ act on a spatial hypersurface
which is spanned by the vectors $\partial_x$, $\partial_y$ and $\partial_z$,
the $h^{00}$ component behaves as a scalar, the $h^{0i}$ as a vector and
the $h^{ij}$ as a 2-rank symmetric tensor, respectively.

\begin{figure}
\includegraphics[width=6cm]{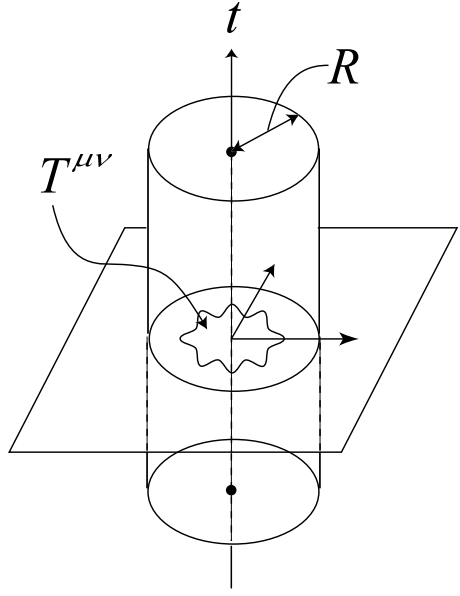}
\caption{\label{fig:tube}The tube and the distribution of the matter}
\end{figure}
Now, let us investigate the solution of the equation (\ref{eq:LinearEin})
under the assumption that the matter exists locally
in the neighborhood of the origin.
Namely, we assume that the matter is confined in the inside of
a (4-dimensional) tube with the radius $R$ and there is no matter
in the outside of the tube (see the figure \ref{fig:tube}).
Additionally, we adopt the boundary condition that $h^{\mu\nu}\to0$ as
$r\to\infty$.
Under these assumptions,
the exterior solutions of the equation (\ref{eq:LinearEin})
is given as follows,
\begin{subequations}\label{eq:SolLE0}
\begin{eqnarray}
h^{00}&=&\frac{4G}{c^4}
    \sum^{\infty}_{l=0}\partial_L
    \left[\frac{M^L(t_{\rm ret})}{r}\right], \\
h^{0i}&=&\frac{4G}{c^4}
    \sum^{\infty}_{l=1}\left\{
    \partial_{L-1}\left[\frac{P^{iL-1}(t_{\rm ret})}{r}\right]
    +\partial_{aL-1}\left[
    \frac{\epsilon^{iam}S^{mL-1}(t_{\rm ret})}{r}\right]\right\}
    +\frac{4G}{c^4}\sum^{\infty}_{l=0}\partial_{aL}
    \left[\frac{\delta^{ia}Q^L(t_{\rm ret})}{r}\right], \\
h^{ij}&=&\frac{4G}{c^4}
    \sum^{\infty}_{l=0}\delta^{ij}\partial_L
    \left[\frac{A^L(t_{\rm ret})}{r}\right]+
    \frac{4G}{c^4}\sum^{\infty}_{l=2}\left\{
        \partial_{L-2}\left[\frac{B^{ijL-2}(t_{\rm ret})}{r}\right]
        +\partial_{aL-2}\left[
        \frac{\epsilon^{am(i}C^{j)mL-2}(t_{\rm ret})}{r}\right]
    \right\} \nonumber \\
&&{~}+\frac{4G}{c^4}\sum^{\infty}_{l=1}\left\{
    \partial_{aL-1}\left[\frac{\delta^{a(i}D^{j)L-1}(t_{\rm ret})}{r}\right]
    +\partial_{abL-1}\left[
    \frac{\delta^{a(i}\epsilon^{j)bm}E^{mL-1}(t_{\rm ret})}{r}
    \right]\right\} \nonumber \\
&&{~}+\frac{4G}{c^4}\sum^{\infty}_{l=0}\partial_{abL}\left[
    \frac{\delta^{ia}\delta^{jb}F^L(t_{\rm ret})}{r}\right],
\end{eqnarray}
\end{subequations}
where we introduce a retarded time $t_{\rm ret}=t-r$.
The $M^L$ {\it etc.} in (\ref{eq:SolLE0}) are the integral constants and
they behave as 3-dimensional tensors.
It should also be noted that
all the 3-dimensional tensors $M^L$ {\it etc.} are perfectly
symmetric and traceless:
$M^{k_1\cdots k_l}=M^{(k_1\cdots k_l)}$ and $M^{ijL}\delta_{ij}=0$ {\it etc}.
We call the $M^L$'s the gravitational multipole moments.
Each gravitational multipole moment is a element of
an irreducible representation space of $SO(3)$
because of its symmetry.

On the other hand, the interior solution is determined by
the distribution of the matter in the tube.
By jointing the interior and the exterior solutions at
the surface of the tube, we can determine
the gravitational multipole moments $M^L$'s.
The important point is that
any object in the exterior gravitational field of the tube cannot distinguish
any interior structure if there is no difference in the $M^L$'s.
In other words, the only $M^L$'s prescribe 
all the external gravitational properties of the matter in the tube.
According to this view,
we can freely deform the distribution of the matter in the inside of the tube
as long as the deformation does not change $M^L$'s.
Among such deformations,
the simplest one is contraction of the matter into a point.
This can be achieved by extrapolating the exterior solution
into the inside of the tube (this is equivalent to $R\to0$).
By this deformation, the extended matter is replaced with
a particle that is represented by a singularity of the gravitational field.
We call this particle as a multipole particle.
Thus, our objective now is to investigate the motion of this
multipole particle. However, before proceeding further,
we examine the properties of the solution (\ref{eq:SolLE0}).
It is because the solution (\ref{eq:SolLE0}) has some gauge freedom
and we have not used any gauge condition in (\ref{eq:SolLE0}) until now.

Firstly, we obtain the following relations among the tensors from
the de Donder condition,
\begin{subequations}
\begin{equation}
\dot{M}^L+P^L+\ddot{Q}^L=0,
\end{equation}
\begin{equation}
\dot{P}^L+B^L+\frac{1}{2}\ddot{D}^L=0,
\end{equation}
\begin{equation}
\dot{S}^L+\frac{1}{2}C^L+\frac{1}{2}\ddot{E}^L=0,
\end{equation}
\begin{equation}
\dot{Q}^L+A^L+\frac{1}{2}D^L+\ddot{F}^L=0,
\end{equation}
\end{subequations}
where dots over tensors show the derivatives by $t$.
Next, since the general form of (\ref{eq:SolLE0})
still includes some gauge freedom,
we can further restrict the form of the solution
by another gauge transformation.
Let us consider the following gauge transformation
\begin{equation}
h^{\mu\nu}_{\rm(new)}=h^{\mu\nu}_{\rm(old)}+{\cal L}_{\xi}
\left\{\sqrt{-g}g^{\mu\nu}\right\},
\end{equation}
where
\begin{subequations}
\begin{eqnarray}
\xi^0&=&\frac{4G}{c^4}\sum^{\infty}_{l=0}\left\{
    \partial_L\left[\frac{Q^L}{r}\right]+
    \frac{1}{2}\partial_L\left[\frac{\dot{F}}{r}\right]\right\} \\
\xi^i&=&\frac{4G}{c^4}\sum^{\infty}_{l=1}\left\{
    \frac{1}{2}\partial_{L-1}\left[\frac{D^{iL-1}}{r}\right]
    +\partial_{aL-1}\left[\frac{\epsilon^{iam}E^{mL-1}}{r}\right]\right\}
    +\frac{4G}{c^4}\sum^{\infty}_{l=0}\frac{1}{2}\partial_{aL}
    \left[\frac{\delta^{ia}F^L}{r}\right]
\end{eqnarray}
\end{subequations}
After this transformation, we become aware that
the tensors $A^L$, $D^L$, $E^L$ and $F^L$ can be set to 0.
Consequently, we obtain the following expression for
a general isolated solution of the linearized Einstein equation,
of which all gauge freedom is fixed:
\begin{subequations}\label{eq:solutionLE}
\begin{eqnarray}
h^{00}&=&\frac{4G}{c^4}\sum^{\infty}_{l=0}\partial_L
    \left[\frac{M^L(t_{\rm ret})}{r}\right] \\
h^{0i}&=&\frac{4G}{c^4}\sum^{\infty}_{l=1}\left\{
    -\partial_{L-1}\left[\frac{\dot{M}^{iL-1}(t_{\rm ret})}{r}\right]
    +\partial_{aL-1}\left[\frac{J^{ia\cdot L-1}(t_{\rm ret})}{r}
    \right]\right\} \\
h^{ij}&=&\frac{4G}{c^4}\sum^{\infty}_{l=2}\left\{
        \partial_{L-2}\left[\frac{\ddot{M}^{ijL-2}(t_{\rm ret})}{r}\right]
        +2\partial_{aL-2}\left[
        \frac{\dot{J}^{a(i\cdot j)L-2}(t_{\rm ret})}{r}
        \right]\right\}
\end{eqnarray}
\end{subequations}
where we define new symbols $J^{ia\cdot L-1}\equiv\epsilon^{iam}S^{mL-1}$.
By these equations, we define the multipole moments of the field created by
the multipole particle. We call the $M^L$ {\it mass multipole moments}
(or mass multipoles), 
and the $J^{ij\cdot L}$ {\it spin multipole moments}
(or spin multipoles).
Note that the mass multipoles begin with mass monopole,
but the spin multipoles begin with spin dipole.
The mass dipole moment, especially, decides
the relation between the center-of-field and
the {\it mean position} of the objects.
If there is any necessity to distinguish $S^L$ and $J^{ij\cdot L}$,
we refer them as {\it original} spin multipoles and
{\it variant} spin multipoles,
respectively. However, we will use the variant spin multipoles in
almost all the case.

\subsection{Multipole moment in oblique frame}
In the last subsection, we obtained
the multipole expansion of the linearized gravitational field
which is based on a representation of $SO(3)$.
However, in the definition of multipole moments,
there is still some freedom of selection of 3-dimensional hyper-surface
which $SO(3)$ acts on. In other words,
"on which 3-dimensional space we define the multipoles?"
Accordingly, what we should consider here is
how we can define multipole moments in an inclined 3-dimensional space
which is not orthogonal to the 4-velocity of the particle.

In order to specify a spatial 3-dimensional subspace, we identify
a timelike vector, namely $v$, which is orthogonal to this subspace.
Simultaneously, we consider $\{e_i\}~(i=1,2,3)$ as unit vectors which span
the hyper-surface. Then, $\{v,e_i\}$ forms an orthogonal tetrad.
We call it $(v,e)$-tetrad frame.
\begin{figure}
\includegraphics[width=5cm]{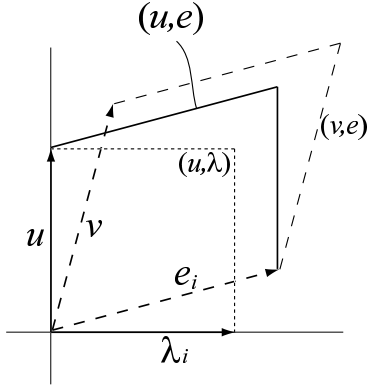}
\caption{\label{fig:frame}Definition of frames.}
\end{figure}
On the other hand, we denote the velocity of the particle by $u$ and
set $\lambda_i=\partial_i~(i=1,2,3)$. 
We call the orthogonal tetrad $\{u,\lambda_i\}$ as $(u,\lambda)$-tetrad frame.
These two tetrad frames, namely $(u,\lambda)$-frame and $(v,e)$-frame,
are related by a Lorenz transformation $\Lambda$ as
\begin{equation}
x^{\mu}=\Lambda^{\mu}{}_{\nu}\bar{x}^{\nu},
\end{equation}
where $x^{\mu}$ is the coordinate of $(u,\lambda)$-frame and
$\bar{x}^{\mu}$ is of $(v,e)$-frame.
Furthermore, let a tetrad constructed by $\{u,e_i\}$ be referred as
$(u,e)$-tetrad frame. This $(u,e)$-frame is the oblique
frame that we should consider (Fig. \ref{fig:frame}).

We cannot apply the similar discussion in the last subsection directly
to the oblique case because the non-orthogonal property
makes the gauge transformation too complicated 
to find an appropriate gauge function $\xi^{\mu}$.
However, we can give general isolated solutions in $(u,e)$-frame as below,
which is almost the same expression (\ref{eq:SolLE0}) in $(u,\lambda)$-frame:
\begin{subequations}\label{eq:SolLE1}
\begin{eqnarray}
\tilde{h}^{00}&=&\frac{4G}{c^4}\sum^{\infty}_{l=0}\tilde{\partial}_L
    \left[\frac{M^L(\tilde{t}_{\rm ret})}{\tilde{r}}\right], \\
\tilde{h}^{0i}&=&\frac{4G}{c^4}\sum^{\infty}_{l=1}\left\{
    \tilde{\partial}_{L-1}\left[\frac{P^{iL-1}(\tilde{t}_{\rm ret})}{\tilde{r}}
    \right]
    +\tilde{\partial}_{aL-1}\left[
    \frac{\epsilon^{iam}S^{mL-1}(\tilde{t}_{\rm ret})}{\tilde{r}}\right]
    \right\}
    +\frac{4G}{c^4}\sum^{\infty}_{l=0}\tilde{\partial}_{aL}
    \left[\frac{\delta^{ia}Q^L(\tilde{t}_{\rm ret})}{\tilde{r}}\right], \\
\tilde{h}^{ij}&=&\frac{4G}{c^4}\sum^{\infty}_{l=0}\delta^{ij}\tilde{\partial}_L
    \left[\frac{A^L(\tilde{t}_{\rm ret})}{\tilde{r}}\right]+
    \frac{4G}{c^4}\sum^{\infty}_{l=2}\left\{
        \tilde{\partial}_{L-2}\left[
        \frac{B^{ijL-2}(\tilde{t}_{\rm ret})}{\tilde{r}}\right]
        +\tilde{\partial}_{aL-2}\left[
        \frac{\epsilon^{am(i}C^{j)mL-2}(\tilde{t}_{\rm ret})}{\tilde{r}}\right]
    \right\} \nonumber \\
&&{~}+\frac{4G}{c^4}\sum^{\infty}_{l=1}\left\{
    \tilde{\partial}_{aL-1}\left[
    \frac{\delta^{a(i}D^{j)L-1}(\tilde{t}_{\rm ret})}{\tilde{r}}\right]
    +\tilde{\partial}_{abL-1}\left[
    \frac{\delta^{a(i}\epsilon^{j)bm}E^{mL-1}(\tilde{t}_{\rm ret})}{\tilde{r}}
    \right]\right\} \nonumber \\
&&{~}+\frac{4G}{c^4}\sum^{\infty}_{l=0}\tilde{\partial}_{abL}\left[
    \frac{\delta^{ia}\delta^{jb}F^L(\tilde{t}_{\rm ret})}{\tilde{r}}\right].
\end{eqnarray}
\end{subequations}
Here $\tilde{h}^{\mu\nu}$ is the components of $h$ in $(u,e)$-frame and
$\tilde{\partial}_a$ is the derivative along the vector $e_a$.
Furthermore, we introduce $\tilde{t}_{\rm ret}$ and $\tilde{r}$
as the retarded time and "oblique" radius, respectively.
If the Lorenz transformation between $(u,\lambda)$ and $(v,e)$
is the boost in $x$-axis direction, namely
$t=\alpha\bar{t}+\beta\bar{x},~x=\beta\bar{t}+\alpha\bar{x},~
(\alpha^2-\beta^2=1)$, then they become
$\tilde{r}=\sqrt{(\alpha\tilde{x})^2+\tilde{y}^2+\tilde{z}^2}$ and
$\tilde{t}_{\rm ret}=\tilde{t}+\beta\tilde{x}-\tilde{r}$,
where $\tilde{x}$ {\it etc.} is the coordinates in $(u,e)$-frame.
However, concrete expression of $\tilde{t}$ and $\tilde{r}$
is not important. The existence of an one-to-one relation
between $\tilde{t}_{\rm ret}$, $\tilde{r}$ and $t_{\rm ret}$, $r$
is essential.

We focus on the fact that both (\ref{eq:SolLE0}) and
(\ref{eq:SolLE1}) are the general isolated solutions.
We can define a one-to-one mapping between these solutions as following,
\begin{eqnarray}
{\cal T}:\left\{
    {\begin{array}{*{20}c}
        h^{\mu \nu}  \\
        t_{\rm ret},r  \\
        \partial_{\mu}   \\
    \end{array}}\right.
    \to
    \left\{
    {\begin{array}{*{20}c}
        {\tilde h^{\mu \nu } }  \\
        {\tilde{t}_{\rm ret},\tilde{r}}  \\
        \tilde{\partial}_{\mu}   \\
    \end{array}} \right. ,
\end{eqnarray}
for example,
\begin{eqnarray}
{\cal T}^{-1}\left(\tilde{h}^{00}\right)
    &=&{\cal T}^{-1}\left(\sum^{\infty}_{l=0}\tilde{\partial}_L
    \left[\frac{M^L(\tilde{t}_{\rm ret})}{\tilde{r}}\right]\right) \nonumber \\
    &=&\sum^{\infty}_{l=0}\partial_L
    \left[\frac{M^L(t_{\rm ret})}{r}\right].
\end{eqnarray}
Given that this mapping is one-to-one, we can construct
a representation of $SO(3)$ on $(u,e)$-frame by using
the representation $\rho$ on $(u,\lambda)$-frame as following,
\begin{equation}\label{eq;representationSO3}
\tilde{\rho}={\cal T}^{-1}\rho{\cal T}.
\end{equation}
According to this construction of the representation,
we can consider that $\tilde{h}^{\mu\nu}$ is also a gauge-fixed solution if
${\cal T}^{-1}(\tilde{h}^{\mu\nu}(\tilde{t}_{\rm ret}))
=h^{\mu\nu}(t_{\rm ret})$ is gauge-fixed.
Therefore, we obtain the general isolated and gauge-fixed solutions in
$(u,e)$-frame under the representation (\ref{eq;representationSO3}),
as following,
\begin{subequations}\label{eq:solutionLE2}
\begin{eqnarray}
\tilde{h}^{00}&=&\frac{4G}{c^4}\sum^{\infty}_{l=0}\tilde{\partial}_L
    \left[\frac{M^L(\tilde{t}_{\rm ret})}{\tilde{r}}\right], \\
\tilde{h}^{0i}&=&\frac{4G}{c^4}\sum^{\infty}_{l=1}\left\{
    -\tilde{\partial}_{L-1}\left[
    \frac{\dot{M}^{iL-1}(\tilde{t}_{\rm ret})}{\tilde{r}}\right]
    +\tilde{\partial}_{aL-1}\left[
    \frac{J^{ia\cdot L-1}(\tilde{t}_{\rm ret})}{\tilde{r}}
    \right]\right\}, \\
\tilde{h}^{ij}&=&\frac{4G}{c^4}\sum^{\infty}_{l=2}\left\{
        \tilde{\partial}_{L-2}\left[
        \frac{\ddot{M}^{ijL-2}(\tilde{t}_{\rm ret})}{\tilde{r}}\right]
        +2\tilde{\partial}_{aL-2}\left[
        \frac{\dot{J}^{a(i\cdot j)L-2}(\tilde{t}_{\rm ret})}{\tilde{r}}
        \right]\right\}.
\end{eqnarray}
\end{subequations}
We define multipoles in the oblique frame by these equations
in the same way of the normal frame case (\ref{eq:solutionLE}).
The important point is that the multipoles in (\ref{eq:solutionLE2})
are perpendicular to $v$ and not $u$,
though those in (\ref{eq:solutionLE})
are perpendicular to $u$.

\subsection{Energy-momentum tensor for multipole particle}
Here, we will derive an energy-momentum tensor for a multipole particle
from the solution of linearized Einstein equation. The relation,
\begin{equation}
\tilde{\Box\mathstrut}\left(\frac{4G}{c^4}
\frac{{\cal M}(\tilde{t}_{\rm ret})}{\tilde{r}}\right)
={}-\frac{16\pi G}{c^4}{\cal M}(\tilde{t})\delta^3(\tilde{r}),
\end{equation}
allows us to derive the following equations from (\ref{eq:solutionLE2}),
\begin{subequations}\label{eq:EM1}
\begin{eqnarray}
\tilde{T}^{00}&=&\sum^{\infty}_{l=0}\tilde{\partial}_L
    \left[M^L(\tilde{t})\delta^3(\tilde{r})\right], \\
\tilde{T}^{0i}&=&\sum^{\infty}_{l=1}\left\{
    -\tilde{\partial}_{L-1}\left[
    \dot{M}^{iL-1}(\tilde{t})\delta^3(\tilde{r})\right]
    +\tilde{\partial}_{aL-1}\left[
    J^{ia\cdot L-1}(\tilde{t})\delta^3(\tilde{r})
    \right]\right\}, \\
\tilde{T}^{ij}&=&\sum^{\infty}_{l=2}\left\{
        \tilde{\partial}_{L-2}\left[
        \ddot{M}^{ijL-2}(\tilde{t})\delta^3(\tilde{r})\right]
        +2\tilde{\partial}_{aL-2}\left[
        \dot{J}^{a(i\cdot j)L-2}(\tilde{t})\delta^3(\tilde{r})
        \right]\right\}.
\end{eqnarray}
\end{subequations}
Here, $\tilde{\Box\mathstrut}$ is the d'Alembertian in $(u,e)$-frame
and we use commutability between $\tilde{\Box\mathstrut}$ and
$\tilde{\partial}_{\mu}$.
This energy-momentum tensor is in the comoving frame of the particle.
By these equations (\ref{eq:EM1}), we define multipole moments of
a particle. In other words,
the multipoles of the particle is defined by the multipoles of
the field created by the particle.  In this definition,
there is no direct relation between
the multipoles and the distributions of the matter which constructs the object.
We regard the point $r=0$ as the position of the particle,
but this position is unrelated to the {\it mean position} of the object.
It is the mass dipole moment to decide the relation between the loci of
the particle and the object. We consider that the particle represents
the object by fixing the mass dipole moment.
Accordingly, we investigate the motion of the particle hereafter.
In order to get the energy-momentum tensor in a general frame,
we only have to transform the coordinates as
\begin{equation}
T^{\mu\nu}=u^{\mu}u^{\nu}\tilde{T}^{00}+
    2u^{(\mu}_{\mathstrut}e^{\nu)}_{i}\tilde{T}^{0i}+
    e_{i}^{\mu}e^{\nu}_{j}\tilde{T}^{ij}.
\end{equation}
After the transformation, we obtain the following expression,
\begin{eqnarray}\label{eq:emsr}
T^{\mu\nu}&=&
\sum^{\infty}_{n=0}\int d\tau \partial_{\rho_1\cdots\rho_n}
    \Bigl(u^{\mu}u^{\nu}
    M^{\rho_1\cdots \rho_n}{\cal D}\Bigr) \nonumber\\
&&{}-\sum^{\infty}_{n=1}\int d\tau \partial_{\rho_1\cdots\rho_{n-1}}
    \Bigl(u^{\mu}\dot{M}^{\nu\rho_1\cdots \rho_{n-1}}{\cal D}\Bigr)
    -\sum^{\infty}_{n=1}\int d\tau \partial_{\rho_1\cdots\rho_{n-1}}
    \Bigl(u^{\nu}\dot{M}^{\mu\rho_1\cdots\rho_{n-1}}{\cal D}\Bigr) \nonumber\\
&&{}+\sum^{\infty}_{n=1} \int d\tau \partial_{\sigma\rho_1\cdots\rho_{n-1}}
    \Bigl(u^{\mu}J^{\nu\sigma\cdot\rho_1\cdots\rho_{n-1}}{\cal D}\Bigr)
    +\sum^{\infty}_{n=1} \int d\tau \partial_{\sigma\rho_1\cdots\rho_{n-1}}
    \Bigl(u^{\nu}J^{\mu\sigma\cdot\rho_1\cdots\rho_{n-1}}{\cal D}\Bigr)
    \nonumber \\
&&{}+\sum^{\infty}_{n=2}\int d\tau \partial_{\rho_1\cdots\rho_{n-2}}
    \Bigl(\ddot{M}^{\mu\nu\rho_1\cdots\rho_{n-2}}{\cal D}\Bigr) \nonumber \\
&&{}+\sum^{\infty}_{n=2}\int d\tau
    \partial_{\sigma\rho_1\cdots\rho_{n-2}}
    \Bigl(\dot{J}^{\sigma\mu\cdot\nu\rho_1\cdots\rho_{n-2}}{\cal D}\Bigr)
    +\sum^{\infty}_{n=2}\int d\tau \partial_{\sigma\rho_1\cdots\rho_{n-2}}
    \Bigl(\dot{J}^{\sigma\nu\cdot\mu\rho_1\cdots\rho_{n-2}}{\cal D}\Bigr),
\end{eqnarray}
where we introduce a covariant delta function
${\cal D}(x)\equiv\frac{\delta^4(x)}{\sqrt{-g(x)}}$ and use a relation,
\begin{equation}
\delta^3(\tilde{r})=\int{\cal D}(\tilde{x}^{\mu}-\tilde{z}^{\mu})d\tau.
\end{equation}
The multipoles in the general frame are defined as follows,
\begin{subequations}
\begin{equation}
M^{\rho_1\cdots\rho_n}=e_{i_1}^{\rho_1}\cdots e_{i_n}^{\rho_n}
    M^{i_1\cdots i_n},
\end{equation}
\begin{equation}
J^{\mu\nu\cdot\rho_1\cdots\rho_n}=
    e_{a}^{\mu}e_{b}^{\nu}\cdot e_{i_1}^{\rho_1}\cdots e_{i_n}^{\rho_n}
    J^{ab\cdot i_1\cdots i_n}.
\end{equation}
\end{subequations}
Noting that the 3-dimensional epsilon $\epsilon^{ijk}$ in
$(u,e)$-frame becomes
$v_{\lambda}\frac{\epsilon^{\lambda\mu\nu\alpha}}{\sqrt{-g}}$
in a general frame, and we obtain the following relation,
\begin{equation}
J^{\mu\nu\cdot\rho_1\cdots\rho_n}=
v_{\lambda}\frac{\epsilon^{\lambda\mu\nu\alpha}}{\sqrt{-g}}
g_{\alpha\beta}S^{\beta\rho_1\cdots\rho_n}.
\end{equation}
Because we have defined the multipoles on the inclined 3-dimensional subspace,
the multipoles must be orthogonal to the normal vector $v$ of the subspace:
\begin{subequations}\label{eq:slicing1}
\begin{equation}
v_{\mu}M^{\mu\cdots}=0,
\end{equation}
\begin{equation}
v_{\rho}J^{\mu\nu\cdot\rho\cdots}=0,~({\rm or~}v_{\mu}S^{\mu\cdots}=0),
\end{equation}
\end{subequations}
and $J^{\mu\nu\cdot\rho\cdots}$ obeys
\begin{equation}\label{eq:slicing2}
v_{\mu}J^{\mu\nu\cdot\rho\cdots}=0
\end{equation}
from the asymmetry of $\epsilon^{\lambda\mu\nu\alpha}$.
The relation for the case of $J^{\mu\nu}$ or $S^{\mu}$ is
known as an spin supplemental condition.

\section{Multipole particle in general relativity}\label{sec:mp in gr}
In this section, we will discuss the energy-momentum tensor
for the multipole particle and its motion on a curved spacetime.

\subsection{Introduce action between particle and gravitational field}
It is easy to obtain an energy-momentum tensor on a curved spacetime
from that given in the last section.
This can be achieved by the "colon-to-semicolon" rule.
Interpreting the rule from the perspective of the gauge theory,
it introduces the interaction between the particle and
the gravitational field by the gauge principle.
However, the way to apply the rule to our case is not unique
because the multipoles {\it etc.} and the covariant derivatives
do not commutate each other.
This issue could be grasped as the identification of a self-field of
the particle on the curved spacetime because the difference by the orders of
them is a combination of the connections and their derivatives.
Therefore, we consider that the order of the multipoles {\it etc.} in
(\ref{eq:emsr}) identifies the self-field and
we will not discuss this problem in detail in this article.
Accordingly, we can give an energy-momentum tensor for
a particle on curved spacetime as following,
\begin{eqnarray}\label{eq:emt}
T^{\mu\nu}&=&
\sum^{\infty}_{n=0}\int d\tau \nabla_{\rho_1\cdots\rho_n}
    \Bigl(u^{\mu}u^{\nu}
    M^{\rho_1\cdots \rho_n}{\cal D}\Bigr) \nonumber\\
&&{}-\sum^{\infty}_{n=1}\int d\tau \nabla_{\rho_1\cdots\rho_{n-1}}
    \Bigl(u^{\mu}\dot{M}^{\nu\rho_1\cdots \rho_{n-1}}{\cal D}\Bigr)
    -\sum^{\infty}_{n=1}\int d\tau \nabla_{\rho_1\cdots\rho_{n-1}}
    \Bigl(u^{\nu}\dot{M}^{\mu\rho_1\cdots\rho_{n-1}}{\cal D}\Bigr) \nonumber\\
&&{}+\sum^{\infty}_{n=1} \int d\tau \nabla_{\sigma\rho_1\cdots\rho_{n-1}}
    \Bigl(u^{\mu}J^{\nu\sigma\cdot\rho_1\cdots\rho_{n-1}}{\cal D}\Bigr)
    +\sum^{\infty}_{n=1} \int d\tau \nabla_{\sigma\rho_1\cdots\rho_{n-1}}
    \Bigl(u^{\nu}J^{\mu\sigma\cdot\rho_1\cdots\rho_{n-1}}{\cal D}\Bigr)
    \nonumber \\
&&{}+\sum^{\infty}_{n=2}\int d\tau \nabla_{\rho_1\cdots\rho_{n-2}}
    \Bigl(\ddot{M}^{\mu\nu\rho_1\cdots\rho_{n-2}}{\cal D}\Bigr) \nonumber \\
&&{}+\sum^{\infty}_{n=2}\int d\tau
    \nabla_{\sigma\rho_1\cdots\rho_{n-2}}
    \Bigl(\dot{J}^{\sigma\mu\cdot\nu\rho_1\cdots\rho_{n-2}}{\cal D}\Bigr)
    +\sum^{\infty}_{n=2}\int d\tau \nabla_{\sigma\rho_1\cdots\rho_{n-2}}
    \Bigl(\dot{J}^{\sigma\nu\cdot\mu\rho_1\cdots\rho_{n-2}}{\cal D}\Bigr).
\end{eqnarray}
Here, the dots over the multipoles show the derivatives by the proper time
$\tau$ of the particle,
\begin{equation}
\dot{~}=\frac{D}{D\tau}\equiv u^{\mu}\nabla_{\mu},
\end{equation}
and the multipoles must also satisfy the slicing conditions
(\ref{eq:slicing1}) and (\ref{eq:slicing2}).
We can obviously obtain the well-known expression for
a point particle from our energy-momentum tensor (\ref{eq:emt})
when we take the only mass monopole moment.

The equation of motion of the particle can be derived from
$T^{\mu\nu}{}_{;\nu}=0$ that is the integrable condition of
the Einstein equation.
However, we give an algorithm using Tulczyjew's theorems
instead of using it directly to obtain the equation of motion.
\newtheorem{tult}{Theorem}
\newtheorem{tull}[tult]{Lemma}
\newtheorem{rem}{Note}
\begin{tull}\label{lemma:1}
For an arbitrary point $x$ in the spacetime and for an arbitrary tensor
$a^{\mu\cdots}$ defined on the world line of a particle,
the following equation holds,
\begin{eqnarray}
&&\int_{-\infty}^{+\infty}
    \nabla_{\alpha}\cdots\nabla_{\beta}\nabla_{\rho}
    \left[a^{\mu\nu\cdots}(\tau)u^{\rho}(\tau){\cal D}(x-z(\tau))\right]
    d\tau \nonumber \\
&&{~~~~~~~~~~~~}=\int_{-\infty}^{+\infty}
    \nabla_{\alpha}\cdots\nabla_{\beta}
    \left[\frac{Da^{\mu\nu\cdots}(\tau)}{D\tau}{\cal D}(x-z(\tau))\right]d\tau,
\end{eqnarray}
where $u^{\mu}$ is 4-velocity of the particle whose position
is denoted by $z(\tau)$.
\end{tull}
\begin{tult}[Tulczyjew]\label{th:tul1}
Let $n^{\mu_1\cdots\mu_k\nu_1\cdots\nu_m}(s)$ be an
arbitrary tensor on a world line and
$N^{\nu_1\cdots\nu_m}$ be a tensor which is
constructed from $n^{\mu_1\cdots\mu_k\nu_1\cdots\nu_m}$ as follows
\begin{equation}
N^{\nu_1\cdots\nu_m}=
    \sum_{k=0}^{n}\int_{-\infty}^{+\infty}
    \nabla_{\mu_1}\cdots\nabla_{\mu_k}
    \left[n^{\mu_1\cdots\mu_k\nu_1\cdots\nu_m}{\cal D}(x-z(\tau))\right]d\tau.
\end{equation}
Then $N^{\nu_1\cdots\nu_m}$ can be reduced to a canonical form
\begin{equation}\label{eq:tulth2-2}
N^{\nu_1\cdots\nu_m}=
    \sum_{k=0}^{n}\int_{-\infty}^{+\infty}
    \nabla_{\mu_1}\cdots\nabla_{\mu_k}
    \left[v^{\mu_1\cdots\mu_k\nu_1\cdots\nu_m}(\tau){\cal D}(x-z(\tau))\right]
    d\tau,
\end{equation}
where
$v^{\mu_1\cdots\mu_k\nu_1\cdots\nu_m}=
v^{(\mu_1\cdots\mu_k)\nu_1\cdots\nu_m}$ and
$u_{\mu_1}v^{\mu_1\cdots\mu_k\nu_1\cdots\nu_m}=0$.
\end{tult}
Remark that it is not necessary for $v^{\mu_1\cdots\mu_k\nu_1\cdots\nu_m}$
to be orthogonal to $u^{\mu}$ by the contraction with the indices $\nu$'s of
$v^{\mu_1\cdots\mu_k\nu_1\cdots\nu_m}$.
Moreover, $v^{\mu_1\cdots\mu_k\nu_1\cdots\nu_m}$ is not
generally equal to $n^{(\mu_1\cdots\mu_k)\nu_1\cdots\nu_m}$.
\begin{tult}[Tulczyjew]\label{th:tul2}
If $N^{\nu_1\cdots\nu_m}$ in theorem \ref{th:tul1} satisfies
\begin{equation}\label{eq:tulth3}
\int_{D}N^{\nu_1\cdots\nu_m}K_{\nu_1\cdots\nu_m}\sqrt{-g}~d^4x=0
\end{equation}
for an arbitrary $K_{\nu_1\cdots\nu_m}$ and 
an arbitrary domain $D$, then all the coefficients 
$v^{\mu_1\cdots\mu_k\nu_1\cdots\nu_m}$ in the canonical form of
$N^{\nu_1\cdots\nu_m}$ must vanish.
\end{tult}
These theorems were given in Tulczyjew's paper.
However, proofs of them were omitted or a little imperfect in
his paper, we give them in the appendix of this article.
Using these theorems, we propose an algorithm as follows:
\begin{enumerate}
\item Calculate the divergence of the Energy-momentum tensor.\label{en:1}
\item Using the lemma \ref{lemma:1} and the commutation relation
    of the covariant derivatives, transform the result of step \ref{en:1}
    into the form in the theorem \ref{th:tul1}.
\item Using the theorem \ref{th:tul2}, derive the equation of motion
    from the coefficients of the canonical form.
\end{enumerate}
We can systematically derive the equations of motion by these steps.

\subsection{spinning particle}
Here, we consider a spinning particle as a non-trivial example.
The spinning particle has angular momentum, and so
we can obtain an energy-momentum tensor for it
when we take $n=0,~1$ in the equation (\ref{eq:emt}):
\begin{equation}\label{eq:TofSpinningParticle}
T^{\mu\nu}=
    \int d\tau\left\{
        \left(Mu^{\mu}u^{\nu}-\dot{M}^{\mu}u^{\nu}-\dot{M}^{\nu}u^{\mu}
            \right){\cal D}
        +\nabla_{\rho}\Bigl[\left(
            M^{\rho}u^{\mu}u^{\nu}+u^{\nu}J^{\mu\rho}+u^{\mu}J^{\nu\rho}
        \right){\cal D}\Bigr]\right\}.
\end{equation}
In this stage, we still leave the mass dipole moment $M^{\mu}$ not to be fixed.
What is to do is to process (\ref{eq:TofSpinningParticle})
according to the algorithm in the last subsection.
For the simplicity, we set
\begin{subequations}\label{eq:mm1}
\begin{eqnarray}
A^{\nu\mu}&\equiv&
    Mu^{\mu}u^{\nu}-\dot{M}^{\mu}u^{\nu}-\dot{M}^{\nu}u^{\mu}, \\
B^{\rho\nu\mu}&\equiv&
    M^{\rho}u^{\mu}u^{\nu}+u^{\nu}J^{\mu\rho}+u^{\mu}J^{\nu\rho}.
\end{eqnarray}
\end{subequations}
Moreover, we decompose $A^{\nu\mu}$ and $B^{\rho\nu\mu}$ into
the compositions parallel and perpendicular to $u$ as follows,
\begin{subequations}
\begin{eqnarray}
A^{\nu\mu}&=&a^{\nu\mu}-u^{\nu}a^{\mu}, \\
B^{\rho\nu\mu}&=&b^{\rho\nu\mu}-u^{\nu}b_1^{\rho^mu}-u^{\rho}b_2^{\nu\mu}
    +u^{\rho}u^{\nu}b^{\mu},
\end{eqnarray}
\end{subequations}
where we define the following variables,
\begin{subequations}\label{eq:mm3}
\begin{eqnarray}
a^{\nu\mu}&\equiv&P^{\nu}{}_{\lambda}A^{\lambda\mu}, \\
a^{\mu}&\equiv&u_{\nu}A^{\nu\mu}, \\
b^{\rho\nu\mu}&\equiv&
    P^{\rho}{}_{\sigma}P^{\nu}{}_{\lambda}B^{\sigma\lambda\mu}, \\
b_1^{\rho\mu}&\equiv&
    P^{\rho}{}_{\sigma}u_{\lambda}B^{\sigma\lambda\mu}, \\
b_2^{\nu\mu}&\equiv&
    u_{\sigma}P^{\nu}{}_{\lambda}B^{\sigma\lambda\mu}, \\
b^{\mu}&\equiv&
    u_{\sigma}u_{\lambda}B^{\sigma\lambda\mu},
\end{eqnarray}
\end{subequations}
and $P^{\mu}{}_{\nu}$ is the projection operator which is defined
by $P^{\mu}{}_{\nu}=\delta^{\mu}{}_{\nu}+u^{\mu}u_{\nu}$.
After some calculation, we obtain the divergence of
the energy-momentum tensor,
\begin{eqnarray}
T^{\mu\nu}{}_{;\nu}&=&
\int d\tau
\Biggl\{
    \nabla_{\nu}\nabla_{\rho}
    \left(
        b^{(\rho\nu)\mu}{\cal D}
    \right)
    +\nabla_{\rho}
    \left[
        \left(
            a^{\rho\mu}+P^{\rho}{}_{\sigma}\frac{D}{D\tau}
            \left(
                u^{\sigma}b^{\mu}-b_1^{\sigma\mu}-b_2^{\sigma\mu}
            \right)
        \right){\cal D}
    \right] \nonumber \\
&&{~~~~~}
    +\left[
        \frac{D}{D\tau}
        \left(
            -a^{\mu}-u_{\sigma}\frac{D}{D\tau}
            \left(
                u^{\sigma}b^{\mu}-b_1^{\sigma\mu}-b_2^{\sigma\mu}
            \right)
        \right)
        +\bigl[\nabla_{\nu},\nabla_{\rho}\bigr]
        \left(
            \frac12b^{\rho\nu\mu}-u^{\nu}b_1^{\rho\mu}
        \right)
    \right]{\cal D}
\Biggr\}.
\end{eqnarray}
Consequently, by using the theorem \ref{th:tul2},
the following equations are acquired:
\begin{subequations}\label{eq:eom0}
\begin{equation}
b^{(\rho\nu)\mu}=0 \label{eq:eom1},
\end{equation}
\begin{equation}
a^{\rho\mu}+P^{\rho}{}_{\sigma}\frac{D}{D\tau}
    \left(u^{\sigma}b^{\mu}-b_1^{\sigma\mu}-b_2^{\sigma\mu}\right)=0
    \label{eq:eom2},
\end{equation}
\begin{equation}
\frac{D}{D\tau}\left(-a^{\mu}-u_{\sigma}\frac{D}{D\tau}
    \left(u^{\sigma}b^{\mu}-b_1^{\sigma\mu}-b_2^{\sigma\mu}\right)\right)
    +\bigl[\nabla_{\nu},\nabla_{\rho}\bigr]
    \left(\frac12b^{\rho\nu\mu}-u^{\nu}b_1^{\rho\mu}\right)=0. \label{eq:eom3}
\end{equation}
\end{subequations}
Let us substitute (\ref{eq:mm1}) and (\ref{eq:mm3}) for (\ref{eq:eom0}).
Firstly, the equation (\ref{eq:eom1}) becomes an identity $0=0$.
Secondly, the equation (\ref{eq:eom2}) turns into
\begin{equation}\label{eq:Ju}
P^{\rho}{}_{\sigma}\left\{-\dot{M}^{\sigma}u^{\mu}
+\frac{D}{D\tau}\left(-J^{\sigma\mu}+M^{\sigma}u^{\mu}\right)\right\}=0.
\end{equation}
By multiplying $u_{\mu}$ to this, we obtain
\begin{equation}\label{eq:Ju0}
\dot{J}^{\sigma\rho}u_{\rho}=0.
\end{equation}
By multiplying $P^{\nu}{}_{\mu}$ to (\ref{eq:Ju}) and
symmetrizing the indices $\nu$ and $\rho$, we get
\begin{equation}\label{eq:ppm}
P^{\nu}{}_{\mu}P^{\rho}{}_{\sigma}\left(
M^{(\sigma}\dot{u}^{\mu)}\right)=0,
\end{equation}
and by antisymmetrizing, we obtain
\begin{equation}\label{eq:asymJM}
P^{\nu}{}_{\mu}P^{\rho}{}_{\sigma}\frac{D}{D\tau}\left(
    -J^{\sigma\mu}+M^{[\sigma}u^{\mu]}\right)=0.
\end{equation}
Thirdly, the equation (\ref{eq:eom3}) with (\ref{eq:Ju0})
changes to
\begin{equation}
\frac{D}{D\tau}\left(Mu^{\mu}-\dot{M}^{\mu}
-\left(u_{\sigma}M^{\sigma}\right)\dot{u}^{\mu}\right)
+\frac12R^{\mu}{}_{\gamma\alpha\beta}u^{\gamma}J^{\alpha\beta}
-R^{\mu}{}_{\alpha\beta\gamma}J^{\alpha\beta}u^{\gamma}
-R^{\mu}{}_{\gamma\alpha\beta}u^{\gamma}M^{\alpha}u^{\gamma}=0.
\end{equation}
Because the relation
$R^{\mu}{}_{\alpha\beta\gamma}+R^{\mu}{}_{\beta\gamma\alpha}
+R^{\mu}{}_{\gamma\alpha\beta}=0$ concludes
\begin{equation}
R^{\mu}{}_{\alpha\beta\gamma}J^{\alpha\beta}u^{\gamma}
=-\frac12R^{\mu}{}_{\gamma\alpha\beta}u^{\gamma}J^{\alpha\beta},
\end{equation}
we can reduce the equation (\ref{eq:eom3}) to
\begin{equation}\label{eq:myeom0}
\frac{D}{D\tau}\left(Mu^{\mu}-\dot{M}^{\mu}
-\left(u_{\sigma}M^{\sigma}\right)\dot{u}^{\mu}\right)
+R^{\mu}{}_{\gamma\alpha\beta}u^{\gamma}\left(J^{\alpha\beta}
-M^{\alpha}u^{\beta}\right)=0.
\end{equation}
Summarizing the result, we obtain the equations
(\ref{eq:Ju0}), (\ref{eq:ppm}), (\ref{eq:asymJM}) and (\ref{eq:myeom0})
as the equation of motion.

Now, we left the mass-dipole $M^{\mu}$ not to be fixed until here.
By choosing a suitable definition of the center-of-mass,
we can show that the equations are equivalent to
the Papapetrou equations,
\begin{subequations}\label{eq:papapetrou}
\begin{equation}
\frac{D}{D\tau}
\left(mu^{\mu}+u_{\sigma}\dot{S}^{\sigma\mu}\right)
+\frac12R^{\mu}{}_{\gamma\alpha\beta}u^{\gamma}S^{\alpha\beta}=0,
\label{eq:papapetrou1}
\end{equation}
\begin{equation}
\dot{S}^{\mu\nu}+\dot{S}^{\mu\sigma}u_{\sigma}u^{\nu}
-\dot{S}^{\nu\sigma}u_{\sigma}u^{\mu}=0.\label{eq:papapetrou2}
\end{equation}
\end{subequations}
To do this, we redefine the mass monopole and spin-dipole as follows,
\begin{subequations}
\begin{equation}
S^{\mu\nu}\equiv2J^{\mu\nu}-M^{\mu}u^{\nu}+M^{\nu}u^{\mu}
\label{eq:SJrelation}
\end{equation}
\begin{equation}
m\equiv M+\left(u_{\sigma}\dot{M}^{\sigma}\right)
\label{eq:mMrelation}
\end{equation}
\end{subequations}
It is obvious that (\ref{eq:asymJM}) becomes
to (\ref{eq:papapetrou2}) under the equation (\ref{eq:SJrelation}).
On the other hand, the replacement (\ref{eq:mMrelation}) makes the contents in
the first parenthesis of (\ref{eq:myeom0}) to
\begin{eqnarray}
Mu^{\mu}-\dot{M}^{\mu}-\left(u_{\sigma}M^{\sigma}\right)\dot{u}^{\mu}
=mu^{\mu}-P^{\mu}{}_{\sigma}\dot{M}^{\sigma}
-\left(u_{\sigma}M^{\sigma}\right)\dot{u}^{\mu}.
\end{eqnarray}
Now, notice the fact that we can set
\begin{equation}
P^{\mu}{}_{\sigma}\dot{M}^{\sigma}+\left(u_{\sigma}M^{\sigma}\right)u^{\mu}
    \equiv\dot{S}^{\mu\sigma}u_{\sigma}, \label{eq:PMSu}
\end{equation}
because of the freedom of $M^{\mu}$.
The mass-dipole $M^{\mu}$ is orthogonal to $v$, so that,
$M^{\mu}$ has three degrees of freedom.
On the other hand, since the equation (\ref{eq:PMSu}) is orthogonal to $u$,
(\ref{eq:PMSu}) means three independent equations.
To sum up, the equation (\ref{eq:PMSu}) is a set of
three ordinary differential equations for
the three independent components of $M^{\mu}$.
According to the theorem on the existence of the solution of
the ordinary differential equations,
we can assure the existence of $M^{\mu}$ satisfying (\ref{eq:PMSu}).
Consequently, we have showed that the equation (\ref{eq:myeom0})
is equivalent to the Papapetrou equation (\ref{eq:papapetrou1}).
Moreover, we can also show the energy-momentum tensor
(\ref{eq:TofSpinningParticle}) is equal to
the energy-momentum tensor given by Tulczyjew \cite{Tulczyjew},
\begin{equation}
T^{\mu\nu}_{\rm Tulczyjew}=
\int d\tau\left\{
    \left(mu^{\mu}u^{\nu}
    -\frac{1}{2}\frac{DS^{\mu\rho}}{D\tau}u_{\rho}u^{\nu}
    -\frac{1}{2}\frac{DS^{\nu\rho}}{D\tau}u_{\rho}u^{\mu}\right){\cal D}
    -\frac{1}{2}\nabla_{\rho}\left[\left(
        S^{\rho\mu}u^{\nu}+S^{\rho\nu}u^{\mu}
    \right){\cal D}\right]
\right\},
\end{equation}
under the replacement (\ref{eq:SJrelation}),
(\ref{eq:mMrelation}) and (\ref{eq:PMSu}).

In the above discussion, we set $M^{\mu}$ to satisfy (\ref{eq:PMSu}).
However, This is not only a unique choice. Rather,
the equations become more simple by setting $M^{\mu}=0$
as well as the case of Newton gravity.
When we set $M^{\mu}$ in this way, we conclude the following equations
\begin{subequations}
\begin{equation}
M\frac{Du^{\mu}}{D\tau}+R^{\mu}{}_{\gamma\alpha\beta}
u^{\gamma}J^{\alpha\beta}=0, \label{eq:myeom1}
\end{equation}
\begin{equation}
\frac{DJ^{\mu\nu}}{D\tau}=0, \label{eq:myeom2}
\end{equation}
\begin{equation}
\frac{DM}{D\tau}=0,\label{eq:myeom3}
\end{equation}
\end{subequations}
as the equation of motion of the spinning particle.
The equation (\ref{eq:myeom1}) was known as
the pole-dipole approximation of the equation of (\ref{eq:papapetrou1}),
but, it is no longer necessary to mention pole-dipole approximation.

\section{summary}\label{sec:summary}
In this article,
we constructed an energy-momentum tensor for a multipole particle,
by considering a gravitational field created by the object.
The multipole moments of the object were defined by
those of the gravitational field, and they were
elements of the irreducible representation space of $SO(3)$.
In this definition of the multipoles, there were two kind of arbitrariness.
One of them is a slicing condition
that decides a spatial hyper-surface, in which
the multipoles are defined. Another one is a dipole condition,
that is concerned with the center-of-field and mean-position of the particle.
By specifying the dipole in a certain manner,
we obtained the Papapetrou equations as the equations of motion.
Another specification led us to more simple representation of
the equations of motion for the spinning particle.
Our equations allow us to understand the motion of the spinning particle
in more simple way.
The deviation from a geodesic can be given by only a coupling term
of the spin of the particle and the curvature of background spacetime,
and the spin evolves by parallel transportation along the world line
of the particle.

The simpler equations bring us the following advantages.
In the case of the Papapetrou equations,
the 4-velocity of the particle is different
from its linear momentum in general.
Therefore, some researchers have imposed the pole-dipole approximation
on the equations of motion. This approximation is based on
the assumption that quadratic terms of the spin have
the same order of quadrupole moment.
However, if we consider a peculiar case that
an object has spin but no quadrupole moment,
for example a spinning {\it rigid body},
the pole-dipole approximation is no longer valid.
Nevertheless, our equations are valid because
the 4-velocity is parallel to the linear momentum in our equation.

\begin{acknowledgments}
The author greatly thanks to Y.Kurita of Kyoto University (GSHES) for
his help in obtaining references.
\end{acknowledgments}

\appendix
\section{Proofs for Tulczyjew's theorems}
\subsection{Proof for lemma \ref{lemma:1}}
Let us focus on the following,
\begin{eqnarray}
\nabla_{\rho}\left(a^{\mu\nu\cdots}(\tau)u^{\rho}{\cal D}\right)
    &=&\frac{\partial}{\partial x^{\rho}}
    \left(a^{\mu\nu\cdots}(\tau)u^{\rho}{\cal D}\right)
    +\Gamma^{\mu}{}_{\rho\sigma}a^{\rho\nu\cdots}u^{\rho}{\cal D}+
    \cdots+\Gamma^{\rho}{}_{\rho\sigma}
    a^{\mu\nu\cdots}u^{\sigma}{\cal D} \nonumber \\
&=&{}-\frac{d}{d\tau}\left(a^{\mu\nu\cdots}{\cal D}\right)
    +\frac{Da^{\mu\nu\cdots}}{D\tau}{\cal D}. \label{eq:a1}
\end{eqnarray}
Note that $a^{\mu\nu\cdot}(\tau)$ and $u^{\mu}{(\tau)}$
are not function of $x$ and they pass through the partial derivatives by $x$.
Additionally, $\frac{d}{d\tau}$ and $\nabla_{\mu}$ are commutate
each other, because $\tau$ is merely a parameter and has no relation to $x$.
Therefore, the first term of the equation (\ref{eq:a1}) becomes
\begin{eqnarray}
    \int_{-\infty}^{+\infty}
    \nabla_{\alpha}\cdots\nabla_{\beta}
    \left[\frac{d}{d\tau}\left(a^{\mu\nu\cdots}{\cal D}\right)\right]d\tau
=\int_{-\infty}^{+\infty}
    \frac{d}{d\tau}\left[
        \nabla_{\alpha}\cdots\nabla_{\beta}
        \left(a^{\mu\nu\cdots}{\cal D}\right)
    \right]d\tau,
\end{eqnarray}
then we can proof the lemma \ref{lemma:1}.

\subsection{Proof for theorem \ref{th:tul1}}
Our proof is based on the mathematical induction.
Firstly, suppose that
$\nabla_{\mu_k}\cdots\nabla_{\mu_1}n^{\mu_1\cdots\mu_k}$,
can be reduced to the following form,
\begin{equation}
\nabla_{\mu_k}\cdots\nabla_{\mu_1}n^{\mu_1\cdots\mu_k}
=\sum_{l=0}^{k}\nabla_{\mu_l}\cdots\nabla_{\mu_1}v^{\mu_1\cdots\mu_l},
\end{equation}
with the perfectly symmetric tensor $v^{\mu_1\cdots\mu_l}$.

Secondly, let us consider a $(k+1)$-rank tensor $n^{\mu_1\cdots\mu_{k+1}}$.
We can suppose that it is symmetry with the indices from  $\mu_1$ to $\mu_k$
due to the assumption of the mathematical induction.
Notice the following equation,
\begin{eqnarray}\label{eq:aaaaa}
n^{(\mu_1\cdots\mu_k)\mu_{k+1}}=
    n^{(\mu_1\cdots\mu_{k+1})}
    +\frac{2}{k+1}\left(
        n^{(\underline{\mu}_1\cdots\mu_k)\underline{\mu}_{k+1}}
        +\cdots+n^{(\mu_1\cdots\underline{\mu}_k)\underline{\mu}_{k+1}}
    \right),
\end{eqnarray}
where we introduce a new notation for anti-symmetrization.
The indices with an underline in (\ref{eq:aaaaa}) means anti-symmetrization
after symmetrization,
for instance, $n^{(\underline{\mu}_1\cdots\mu_k)\underline{\mu}_{k+1}}
=\frac12\left(n^{(\mu_1\cdots\mu_k)\mu_{k+1}}-n^{(\mu_{k+1}\cdots\mu_k)\mu_1}
\right)$. Using this relation (\ref{eq:aaaaa}), we obtain
\begin{eqnarray}\label{eq:apth2}
\nabla_{\mu_{k+1}}&\cdots&\nabla_{\mu_1}n^{(\mu_1\cdots\mu_k)\mu_{k+1}}
    =\nabla_{\mu_{k+1}}\cdots\nabla_{\mu_1}n^{(\mu_1\cdots\mu_{k+1})}
    \\ \nonumber
&&+\frac{1}{k+1}\left(
    [\nabla_{\mu_{k+1}},\nabla_{\mu_1}]\nabla_{\mu_{k}}\cdots\nabla_{\mu_2}
    +\cdots+[\nabla_{\mu_{k+1}},\nabla_{\mu_k}]
    \nabla_{\mu_{k-1}}\cdots\nabla_{\mu_1}
    \right)n^{(\mu_1\cdots\mu_k)\mu_{k+1}}.
\end{eqnarray}
The terms in the parenthesis of the right hand side of
the equation (\ref{eq:apth2}) can be written into the form of
the sum of $(R^{\alpha}{}_{\beta\mu\nu}{\rm~or~}
\nabla\cdots\nabla R^{\alpha}{}_{\beta\mu\nu})\times n^{\rho\cdots}$.
We can apply the assumption of the induction to these terms.
Therefore, we have proofed
\begin{equation}
\nabla_{\mu_{k+1}}\cdots\nabla_{\mu_1}n^{\mu_1\cdots\mu_{k+1}}=
    \sum_{l=0}^{k+1}\nabla_{\mu_l}\cdots\nabla_{\mu_1}v^{\mu_1\cdots\mu_l}.
\end{equation}

Next, let us proof the orthogonality of $v^{\mu_1\cdots\mu_l}$ and $u^{\mu}$.
Define a projection operator
$P^{\mu}{}_{\nu}=\delta^{\mu}{}_{\nu}+u^{\mu}u_{\nu}$.
This is orthogonal to $u^{\mu}$: $P^{\mu}{}_{\nu}u^{\nu}=0$.
We can write an arbitrary tensor $X^{\mu_1\cdots}$ in the form
$X^{\mu_1\cdots}=P^{\mu_1}{}_{\nu_1}\cdots X^{\nu_1\cdots}+
\cdots+u^{\mu_1}\left(u_{\nu_1}X^{\nu_1\cdots}\right)+\cdots$.
Redefine $v^{\mu_1\cdots\mu_l}$ by the tensor projected by $P^{\mu}{}_{\nu}$
instead of old $v^{\mu_1\cdots\mu_l}$, then
we can assure of the orthogonality of
new $v^{\mu_1\cdots\mu_l}$ and $u^{\mu}$.

\subsection{Proof for theorem \ref{th:tul2}}
Let us denote the two intersection points of the domain $D$ and
the world line of the particle by $z(p)$ and $z(q)$ $(p<q)$.
At the end points $z(p)$ and $z(q)$,
we can choose the values of $K_{\nu_1\cdots\nu_m}$ and
its derivatives lower rank than $n-1$ to vanish for the sake of arbitrariness
of $K_{\nu_1\cdots\nu_m}$. Then, substituting (\ref{eq:tulth2-2}) for
(\ref{eq:tulth3}) and integrating it by parts repeatedly,
we obtain the following equation,
\begin{equation}\label{eq:aptul3-1}
\sum_{k=0}^{n}\int_p^qd\tau(-1)^k
\bigl[\nabla_{\mu_k}\cdots\nabla_{\mu_1}K_{\nu_1\cdots\nu_m}
\bigr]_{x=z(\tau)}v^{\mu_1\cdots\mu_k\nu_1\cdots\nu_m}=0.
\end{equation}
Because the values of $K_{\nu_1\cdots\nu_m}$ at $z(p)$ and $z(q)$ are
fixed, we can not assume the behavior of $K_{\nu_1\cdots\nu_m}$
along the world line of the particle. However,
there is still freedom to fix the behaviors of
$K_{\nu_1\cdots\nu_m}$ in the direction perpendicular to the world line.
Therefore, we can set its spatial derivatives to vanish on the world line.
In addition, we change the frame to the comoving frame of the particle.
Consequently, (\ref{eq:aptul3-1}) becomes
\begin{equation}\label{eq:aptul3-2}
\int_p^qd\tau(-1)^n
\bigl[\partial_{i_n}\cdots\partial_{i_1}K_{\nu_1\cdots\nu_m}
\bigr]_{x=z(\tau)}v^{i_1\cdots i_n\nu_1\cdots\nu_m}=0.
\end{equation}
Note that all the derivatives are only spatial ones
because of $u_{\mu_1}v^{\mu_1\cdots\mu_n\nu_1\cdots\nu_m}=0$,
and also note that the indices $i_1$ to $i_n$ are fully symmetry because of
commutability of the partial derivatives.
Now, there remains the arbitrariness of the value of
$\bigl[\partial_{i_n}\cdots\partial_{i_1}
K_{\nu_1\cdots\nu_m}\bigr]_{x=z(\tau)}$,
so that, $v^{i_1\cdots i_n\nu_1\cdots\nu_m}=0$ is concluded
in order for (\ref{eq:aptul3-2}) to be valid.
This means
\begin{equation}
v^{\mu_1\cdots\mu_n\nu_1\cdots\nu_m}=0,
\end{equation}
in general coordinates.
Repeating the same discussion, it is concluded that
all coefficients $v^{\mu_1\cdots\nu_1\cdots}$ must be equal to $0$.


\end{document}